\documentclass[twocolumn,showpacs,nofootinbib,preprintnumbers,amsmath,amssymb,superscriptaddress]{revtex4}

\usepackage{graphicx}
\usepackage{dcolumn}
\usepackage{bm}

\def\lsim{\mathrel{\raise.3ex\hbox{$<$\kern-.75em\lower1ex\hbox{$\sim$}}}}
\def\gsim{\mathrel{\raise.3ex\hbox{$>$\kern-.75em\lower1ex\hbox{$\sim$}}}}

\begin{document}


\title{Distinguishing Between Dark Matter and Pulsar Origins of the ATIC Electron Spectrum With Atmospheric Cherenkov Telescopes}

\author{Jeter Hall}
 \email{jeter@fnal.gov}
\affiliation{Fermi Center for Particle Astrophysics, Fermi National Accelerator Laboratory, Batavia, USA}
\author{Dan Hooper}
\email{hooper@fnal.gov}
\affiliation{Theoretical Astrophysics, Fermi National Accelerator Laboratory, Batavia, USA}
\affiliation{Department of Astronomy and Astrophysics, The University of Chicago, USA}

\date{\today}

\begin{abstract}
Recent results from the Advanced Thin Ionization Calorimeter 
(ATIC) balloon experiment have identified the presence of a spectral feature between approximately 300 and 800 GeV in the cosmic ray electron spectrum. This spectral feature appears to imply the existence of a local ($\lsim 1$ kpc) source of high energy electrons. Emission from a local pulsar and dark matter annihilations have each been put forth as possible origins of this signal. In this letter, we consider the sensitivity of
ground based atmospheric Cherenkov telescopes to electrons and show that observatories such as HESS and VERITAS should
be able to resolve this feature with sufficient precision to discriminate between the dark matter and pulsar hypotheses with considerably greater than 5$\sigma$ significance, even for conservative assumptions regarding their performance. In addition, this feature provides an opportunity to perform an absolute calibration of
the energy scale of ground based, gamma ray telescopes.
\end{abstract}

\pacs{95.55.Ka; 98.70.Rz; 95.35.+d; FERMILAB-PUB-08-528-A}
\maketitle


The ATIC balloon experiment has recently published data revealing a feature in the cosmic ray electron (plus positron) spectrum between approximately 300 and 800 GeV, peaking at around 600 GeV~\cite{atic}. Additionally, the PAMELA collaboration has reported an anomalous rise in the cosmic ray
positron fraction (the positron to positron-plus-electron ratio) above 10 GeV~\cite{pamela}, confirming earlier indications from HEAT~\cite{heat} and AMS-01~\cite{ams01}.  These observations suggest the presence of a relatively local source or sources of energetic cosmic ray electrons and positrons.  Furthermore, the WMAP experiment has revealed an excess of microwave emission from the central region of the Milky Way which has been interpreted as synchrotron emission from a population of electrons/positrons with a hard spectral index~\cite{haze}. Taken together, these observations suggest that energetic electrons and positrons are surprisingly ubiquitous throughout our galaxy.

Although the origin of these electrons and positrons is not currently known, interpretations of the observations have focused two possibilities: emission from pulsars~\cite{pulsars,pulsars2}, and dark matter annihilations~\cite{darkmatter,darkmatter2}. In order for dark matter annihilations throughout in the Milky Way halo to produce a spectrum with a shape similar to that observed by PAMELA and ATIC, however, a large fraction of the annihilations must proceed to electron-positron pairs, or possibly $\mu^+ \mu^-$ or $\tau^+ \tau^-$~\cite{darkmatter}.  Dark matter annihilations to $e^+ e^-$ also result in a distinctive feature in the cosmic ray electron spectrum. In particular, they are predicted to produce an edge in the electron spectrum that drops off suddenly at $E_e = m_{\rm DM}$. In contrast, pulsars and other astrophysical sources of cosmic ray electrons are expected to produce spectra which fall off more gradually. Although the current data from ATIC are not detailed enough to discriminate between a feature with a sudden edge (dark matter-like) or graduate cutoff (pulsar-like), such a discrimination could become possible if the electron spectrum were to be measured with greater precision.

In 2004, Baltz and Hooper suggested that atmospheric Cherenkov telescopes (ACTs) would have the ability to identify sudden features in the cosmic ray electron spectrum, such as that resulting from dark matter annihilations to electron-positron pairs~\cite{baltz}. In this letter, we revisit this possibility and show that ground based, gamma ray telescopes should be capable of resolving the spectral feature seen by ATIC with much greater precision than is currently available. We find that if the ATIC excess is the result of dark matter annihilations to $e^+ e^-$, the edge-like feature will be unmistakable to these telescopes.



The ATIC experiment consists of a silicon matrix detector and an array of Bismuth Germanate crystals, designed to measure the
charge magnitude and energy of incident cosmic rays, respectively.  The geometrical collecting area
of the calorimeter is $\sim0.5$ m$^{2}$.  There have been two flights so
far in the ATIC program: one over two and one half weeks near the end of $2000$, and another during the 2007-2008 Antarctic summer.

In agreement with previous experiments~\cite{previous}, the electron spectrum between 20 GeV and a few hundred GeV has been measured by ATIC to take the form of a steeply falling power law
spectrum, $dN_e/dE_e \propto E_e^{-\alpha}$, with $\alpha \approx 3.2$.  Above a few hundred GeV, however, ATIC observes a significant hardening of the spectrum. This behavior continues up to $\sim$600 GeV, at which point the spectrum falls rapidly.

The energy of this spectral feature tells us something about the proximity of the responsible source(s). As they propagate through the galaxy, cosmic ray electrons lose energy via inverse Compton scattering and synchrotron processes.  For reasonable estimates of the radiation and magnetic field densities in the local Milky Way, a 600 GeV electron is expected to lose energy at a rate of approximately $\sim$1 MeV per year. These energy losses, combined with the rate of diffusion, require the source(s) of the ATIC feature to be no further than a few kiloparsecs from the Solar System.

\begin{figure}
\includegraphics[width=250 pt]{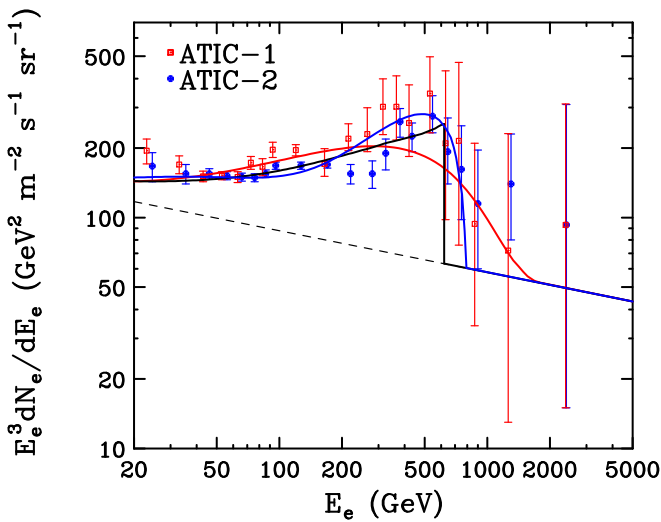}
\caption{\label{atic_fit} The cosmic ray electron spectrum as measured by ATIC~\cite{atic} compared to the spectrum predicted from three possible sources: a nearby pulsar (red), annihilation of $800$ GeV dark matter annihilating to $W^+W^-$ (blue), and annihilation of $620$ GeV Kaluza-Klein dark matter (which annihilates to  $e^+ e^-$, $\mu^+ \mu^-$, and $\tau^+ \tau^-$ 20\% of the time each). In each case, the source spectrum was added to a background power-law spectrum with a spectral slope of -3.2 (dashed).}
\end{figure}

In Fig.~\ref{atic_fit}, we show the electron spectrum predicted for three scenarios, each of which provide a reasonably good fit to the ATIC data (which are also shown). The black line denotes the case of a 620 GeV Kaluza-Klein dark matter particle in a model with a universal extra dimension~\cite{ued}. The most phenomenologically important aspect of dark matter in this model is that it annihilates to $e^+ e^-$, $\mu^+ \mu^-$, and $\tau^+ \tau^-$ 20\% of the time each, leading to a very hard electron spectrum~\cite{Hooper:2004xn}. The annihilations to $e^+ e^-$ lead to the sudden drop in the spectrum at 620 GeV, which is clearly seen in the figure. We have normalized the rate of dark matter annihilation to best fit the ATIC data (which requires a large boost factor).

In contrast, the red line represents the case of emission from a nearby pulsar. Following Ref.~\cite{pulsars}, we consider the pulsar B0656+14 which is 290 parsecs away from the Solar System and 110,000 years old. To be compatible with the spectrum observed by ATIC, we have used an injected spectrum (at source) of the form $dN_e/dE_e \propto E_e^{-1.5} \exp(-E_e/600\, {\rm GeV})$, and a total energy output in $e^+ e^-$ pairs of $3\times 10^{48}$ erg. Although this exponential form leads to a suppression of the electron spectrum at high energies, the cutoff it not nearly as sudden as in the case of dark matter annihilating to pairs. 

As a third case, we show in blue the result for an 800 GeV dark matter particle annihilating to $W^+ W^-$. In contrast to the previous scenarios, here we have adopted boundary conditions for the diffusion region at 1 kiloparsec above and below the Galactic Plane (4 kpc was used in the other cases). This choice suppresses the lower energy range of the spectrum, which otherwise would have exceeded the ATIC measurements. It also should be noted that dark matter annihilations to $W^+ W^-$ distributed throughout the halo of the Milky Way are expected to exceed the gamma ray and antiproton fluxes observed if the annihilation rate is normalized to the electron/positron signals observed by ATIC or PAMELA~\cite{gammaantiproton}. With this in mind, we include this case for comparison.

Among these three scenarios, the most distinguishing feature is the sudden cutoff predicted for dark matter annihilations to $e^+ e^-$ (and to a lesser degree, the nearly as sudden cutoff in the dark matter to $W^+ W^-$ case).  Traditional, bottom-up 
sources of high energy radiation, such as supernova remnants or accreting compact objects,
have an intrinsically softer spectrum and a considerably more gentle cutoff.  Dark matter annihilation, especially in the case of direct annihilation
to leptons, can lead to a significantly harder spectrum with an associated abrupt cutoff
at the dark matter particle's mass.  The spectral shape at lower energies is dominated by details of the diffusion model, so is not as sensitive to the nature of the source.

Unfortunately, current cosmic ray balloon and satellite experiments simply do not have the exposure required to measure the cosmic ray electron spectrum above a few hundred GeV with high precision. For greater exposure and higher precision, we have to turn to ground based experiments. Instead of observing a given cosmic ray or gamma ray directly, ground based experiments are designed to observe secondary particles produced in the resulting air shower.  At a few hundred GeV, the maximum of shower development occurs at approximately 10 kilometers above the Earth, and thus the only particles to reach ground level are photons, muons,
and neutrinos.  The brightest component of these showers is the Cherenkov radiation. Although atmospheric Cherenkov telescopes are typically used as gamma ray detectors, they are also capable and well suited for studying the cosmic ray electron spectrum over the energy range of hundreds of GeV to a few TeV.


\begin{figure}
\includegraphics[width=250 pt]{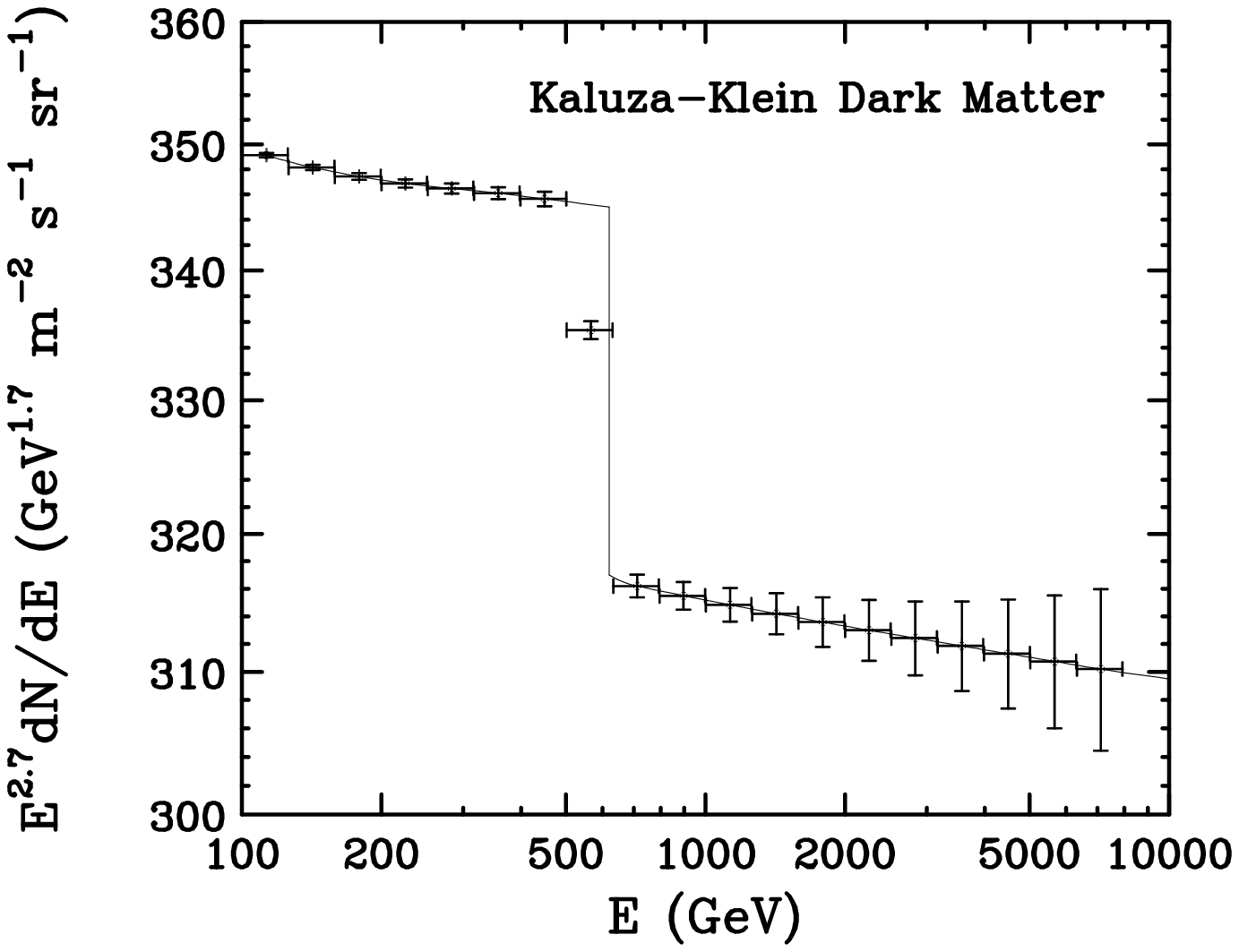}\\
\includegraphics[width=250 pt]{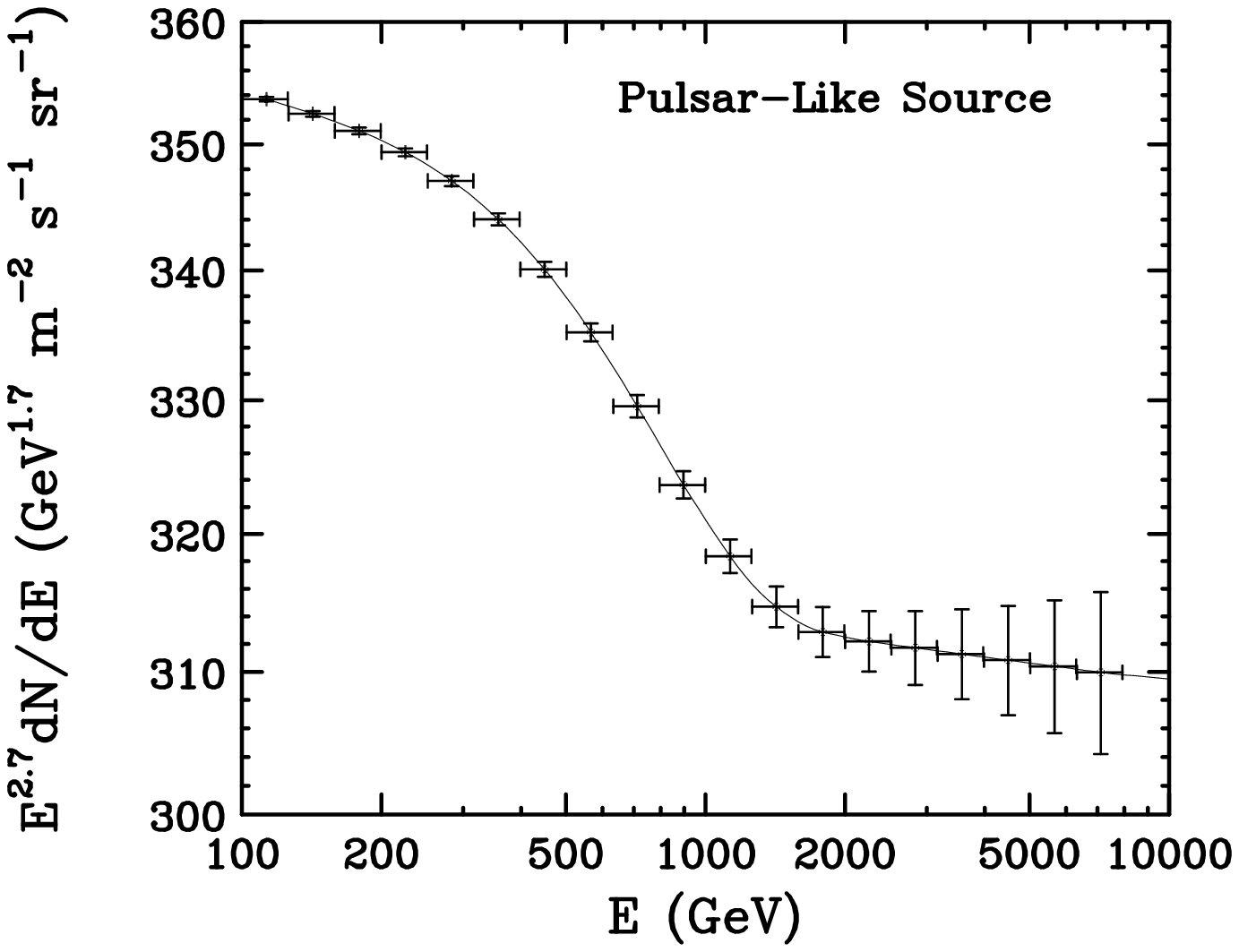}
\caption{\label{act_spectrum} 
The projected cosmic ray electron spectrum for an ACT such as HESS or VERITAS (including misidentified hadrons). Errors are statistical. The very large collecting area of ACTs allow for a much more detailed measurement than is possible from balloon or satellite experiments. With such a measurement the two scenarios shown above are clearly distinguishable.}
\end{figure}

Atmospheric Cherenkov telescopes, including HESS~\cite{hess}, VERITAS~\cite{veritas}, and MAGIC~\cite{magic}, have recently improved their sensitivities
to the point of detecting many classes of galactic and
extragalactic gamma ray sources.  These telescopes have established the
energy regime between $100$ GeV and $100$ TeV as an important window into nature's most powerful
particle accelerators.  Shell-type supernovae, pulsars, binary systems, and active galactic
nuclei have each been established as sources of high energy particles.

ACTs can detect showers induced by gamma rays, cosmic ray electrons, and hadronic cosmic rays. Based on the morphology of such showers, it is possible to efficiently identify and reject approximately $99\%$ of the hadronic cosmic rays, which constitute the dominant
component of the cosmic ray flux and the majority of a typical ACT's data set.
%
%
Although air showers from electron and gamma ray primaries have identical
morphologies and, therefore, currently cannot be distinguished on an event-by-event basis, the anisotropic nature of the majority of gamma rays can be used to identify and reject any large gamma ray component. In effect, the cosmic ray electron spectrum measured by an ACT consists of a combination of electrons, misidentified hadrons, and a smaller number of diffuse gamma rays. The ability of ACTs to make a clean measurement of the electron spectrum thus depends critically on their ability to consistently reject showers initiated by hadronic cosmic rays. Methods are currently being investigated which could increase the proton rejection to $99.9\%$, leading to a nearly clean cosmic ray electron sample at energies where traditional balloon and satellite measurements are exposure limited. To be conservative, however, we assume that only 99\% of the hadronic primaries will be rejected.

Since ACTs use the atmosphere as a calorimeter, their effective 
collecting areas ($\sim 10^5$ m$^2$) are much larger than satellite and balloon based experiments ($\sim 1$ m$^2$).
Although the systematic errors due to calibrating the 
response of the atmosphere and detector across a number of fields-of-view are beyond
the scope of this paper, it appears that such systematics are understood at a level which will allow for a detailed study of the cosmic ray electron spectrum with an energy resolution of $\sim$15\%~\cite{resolution}, including over the energy range containing the feature observed by the ATIC experiment.

To assess the prospects for an experiment such as HESS or VERITAS to resolve the spectrum of the ATIC feature, we have simulated a number of experiments using realistic exposures and rejection levels.  A typical ACT collects $1000$
hours of exposure each year over a lifetime of at least $4$ years.  Excluding surveys of the Galactic Plane, most of the data include few diffuse gamma rays. Observations of the fields-of-view surrounding most extragalactic objects are, in particular, expected to contain a low levels of diffuse gamma rays, and thus will be especially useful for studying cosmic ray electrons.  We assume that $\sim50\%$ of the total ACT exposure is useful for measuring the cosmic ray electron spectrum. We take a typical ACT field-of-view, $\Delta\Omega \sim 0.002$, and a collecting area of $2\times10^5$ m$^2$.

In Fig.~\ref{act_spectrum}, we show the projected ACT statistical errors 
for the case of Kaluza-Klein dark matter (top) and a nearby pulsar (bottom), each with spectra as shown in Fig.~\ref{atic_fit}. These (projected) measured spectra contain both cosmic ray electrons and misidentified hadrons. Although the one percent of proton initiated air showers overwhelms the electron signal by more than an order of magnitude, it does not
substantially distort the electron spectral features that have been suggested to fit
the ATIC spectrum.
The statistical power resulting from the large collecting area of the ACT is apparent when compared to the much larger error bars shown in Fig.~\ref{atic_fit}.

To distinguish between a dark matter and pulsar origin of the ATIC feature, we focus on the energy range around the feature itself (at lower energies, the spectrum depends strongly on the properties of the diffusion model). In the pulsar case, we project that the spectrum could be measured by an ACT to fall with an average slope of $dN/dE \propto E^{-2.7+\beta}$, $\beta=-0.0676 \pm 0.0065$, between 400 and 800 GeV. The same measurement in the case of Kaluza-Klein dark matter would yield a value of $\beta=-0.1845 \pm 0.0065$. With such small statistical errors, the measurement from an ACT would enable these two possibilities to be distinguished at approximately the 18$\sigma$ level. Although systematic errors will reduce this significance somewhat, we expect such a measurement to be capable of distinguishing between these possibilities with very high significance.



As an additional point, it has been difficult to get an absolute calibration of ACT experiments.  All results base their energy scales on simulations and have systematic
errors at the level of $20\%$--$30\%$.  There is a planned calibration of
atmospheric Cherenkov telescopes with the Fermi gamma ray space telescope, but
the overlapping energy range is limited to the lowest energies observable by ACTs.  The ATIC experiment was calibrated at a
particle beam in CERN and thus can be used to provide a valuable energy calibration for ACT experiments.

In summary, the large collecting areas of ACTs provide them with the statistical power required to investigate the feature in the cosmic ray electron
spectrum reported by the ATIC experiment. Existing data from the HESS or VERITAS experiments should be capable of resolving the spectrum of this feature in detail, and distinguish between dark matter and pulsar origins of the observed high energy cosmic ray electrons with very high statistical significance. Additionally, this feature could also lead to an absolute calibration of the ACT energy scale.

\begin{acknowledgments}

We would like to acknowledge Ted Baltz for suggesting that ACTs could resolve the cutoff in the electron spectrum predicted for selected dark matter candidates.  Everything is a little clearer after a few pints of Scottish ale! DH is supported by the Fermi Research Alliance, LLC
  under Contract No.~DE-AC02-07CH11359 with the US Department of
  Energy and by NASA grant NNX08AH34G.
\end{acknowledgments}

\end{document}